\documentclass{article}

\usepackage{amssymb}
\usepackage{amsmath}

\newtheorem{thm}{Theorem}[section]

\newtheorem{rem}[thm]{Remark}

\newcommand{\Left}{{\cal H}}

\begin{document}

\title{Location of the Adsorption Transition for Lattice Polymers}
\author{Neal Madras \\ Department of Mathematics and Statistics \\
York University \\ 4700 Keele Street  \\ Toronto, Ontario  M3J 1P3 Canada 
\\  {\tt  madras@mathstat.yorku.ca} }
\maketitle

\begin{abstract} 
We consider various lattice models of polymers:  
lattice trees, lattice animals, and self-avoiding walks.  The polymer interacts 
with a surface (hyperplane), receiving a unit energy reward for each site in the 
surface.  There is an adsorption transition of the polymer at a critical value of $\beta$, the inverse temperature.
We present a new proof of the result of Hammersley, Torrie, and Whittington (1982)
that the transition occurs at a strictly positive value of $\beta$
when the surface is impenetrable,
i.e.\ when the polymer is restricted to a half-space.  In contrast, for a
penetrable surface, it is an open problem to prove that the transition occurs at $\beta=0$
(i.e., infinite temperature).  We reduce this problem to showing that the 
fraction of $N$-site polymers whose span is less than $N/\log^2 N$ is 
not too small.
\end{abstract}


\section{Introduction}
  \label{sec.intro}

We shall work in the $d$-dimensional hypercubic lattice $\mathbb{L}^d$ ($d\geq 2$), 
with sites $x=(x_1,\ldots,x_d)\in \mathbb{Z}^d$ and edges connecting nearest neighbours.  
Let $\mathbb{L}^d_+$ be the part of $\mathbb{L}^d$ in the half-space $x_1\geq 0$.

Here is our ``big picture'' of adsorption for lattice polymer models.  We have 
a surface in our space $\mathbb{L}^d$ (in our case, the hyperplane $x_1=0$).
For each $N\geq 1$, we have a finite set ${\cal P}_N$ of possible configurations of a 
polymer molecule of size $N$ 
attached to a fixed site in the surface (the origin).  In this paper, ${\cal P}_N$ will be the 
set of lattice trees or lattice animals (representing branched polymers) or self-avoiding 
walks (representing linear polymers) with $N$ sites (representing monomers).
These are classical lattice models of polymer configurations (see for example
de Gennes 1979 and Vanderzande 1998). 
Each polymer $\rho$ is rewarded according to the number $\sigma(\rho)$ 
of sites of $\rho$ that lie in the surface.  For real $\beta$, we define the partition function
\begin{equation}
   \label{eq.Zgen}
   Z_N(\beta)   \; :=\;  \sum_{\rho\in {\cal P_N} } \exp(\beta \sigma(\rho))  \,.
\end{equation}
The absolute value of $\beta$ represents the inverse temperature; the sign of $\beta$ tells
us whether the surface is attractive or repulsive.  In our cases, there exists a 
\textit{limiting free energy}
\begin{equation}
   \label{eq.Fgen}
      {\cal F}(\beta)  \;:=\;  \lim_{N\rightarrow\infty}\frac{1}{N}\,\log Z_N(\beta) \,.
\end{equation}
The limit  ${\cal F}(\beta)$ is a finite non-decreasing function of $\beta$
that is automatically convex (e.g.\ Lemma 4.1.2 of Madras and Slade)
and hence continuous.

In particular, we have $\lim_{N\rightarrow\infty}|{\cal P}_N|^{1/N}\,=\, \exp ({\cal F}(0))$
(where the cardinality of a set $A$ is denoted $|A|$).
In our models, we also find that ${\cal F}(\beta)\,=\,{\cal F}(0)$ for every negative $\beta$,
which says that in the repulsive regime, the energy imparted by surface interaction is 
negligible---i.e., the polymer desorbs and most of it does not lie in the surface.
We say that $\{\beta: {\cal F}(\beta)\,=\,{\cal F}(0)\}$ is the \textit{desorbed} regime,
and $\{\beta: {\cal F}(\beta)\,>\,{\cal F}(0)\}$ is the \textit{adsorbed} regime.
There is an \textit{adsorption transition} at the critical point $\beta_c$ which is the right
endpoint of the desorbed regime. We know that $\beta_c$ is finite (Hammersley, Torrie and Whittington, 1982).

In the context of polymer modelling, the surface could either be
impenetrable (e.g., the wall of a container) or penetrable (e.g., an interfacial
layer between two fluids).   We shall always represent the surface by the 
hyperplane $x_1=0$.  In the impenetrable case, the polymer configurations 
will be restricted to the half-space $\mathbb{L}^d_+$.  We shall write
$\beta_c^+$ and $\beta_c^P$ to denote the adsorption critical points for the 
impenetrable and penetrable models respectively.

A basic qualitative question about the adsorption transition
is whether $\beta_c$ is zero or nonzero---i.e., whether the transition occurs at infinite
or at finite temperature.  It turns out that when the surface is impenetrable, 
then $\beta_c^+>0$.  This had been proven
by other authors (Hammersley et al.\ 1982, for self-avoiding walks;
Janse van Rensburg and You, 1998, for lattice trees), 
but we present  a new and shorter proof. 
In the case of a penetrable surface, with the polymers not restricted to a half-space,
 it is generally believed that $\beta_c^P=0$.  It is an open problem to prove this
rigorously.  We do not fully solve this problem, but we show that it is a rigorous consequence of
a weak assertion about the diameter of polymers which seems to be beyond reasonable doubt.
Specifically, let  the span of the polymer $\rho$ be the maximum value of $|u_1-v_1|$
where $u$ and $v$ range over all sites of $\rho$.  Let $f_N$ be the 
fraction of polymers in ${\cal P}_N$ whose span is at most $N/\log^2N$.  We prove that if 
$f_N$ is bounded below $N^{-\delta}$ for some fixed $\delta$, then $\beta_c$ must be 
zero.  This condition is much weaker than the standard scaling assumption about
polymers, which is that the average span of members of ${\cal P}_N$ scales
as $N^{\nu}$ for some $\nu<1$.

It is worth remarking that the methods of Hammersley et al.\ (1982) and Janse van Rensburg and 
You (1998) yields an explicit positive lower bound on $\beta^+_c-\beta^P_c$; the strict 
positivity of $\beta^+_c$ is then a corollary of this result and 
the relatively easy observation that $\beta_c^P\geq 0$.  In contrast, the method of the
present paper provides an explicit positive lower bound on $\beta^+_c$ but does not
give a direct proof that $\beta^+_c>\beta^P_c$.

Beaton et al.\ (2014) considered the important special case of self-avoiding walks 
on the hexagonal lattice, and proved that $\beta^{+}_c=\ln (1+\sqrt{2})$, thus verifying
a prediction of Batchelor and Yung (1995).  This result depends on special properties
of the hexagonal lattice, and seems difficult to generalize.

We note that when ${\cal P}_N$ is the set of $N$-step nearest-neighbour random walk paths 
(not necessarily self-avoiding), then a relatively straightforward application of generating
functions shows that $\beta_c$ is 0 in the penetrable case and is strictly positive (in fact 
equal to $\ln(2d/(2d-1))$) in the impenetrable case (see for example Hammersley, 1982).
The book of Giacomon (2007) deals extensively with related random walk models.

Our proofs are simplest in the case of lattice trees and lattice animals.  The same methods
work for self-avoiding walks, but some technical modifications are necessary.

Here is the organization of the rest of the paper.  The results are stated formally in Section 
\ref{sec.results}.  After Section \ref{sec.defs} sets up the basic framework and some terminology,
Sections \ref{sec.BP1} and \ref{sec.SAW1} present the results for lattice trees (and lattice 
animals) and for self-avoiding walks respectively.  
Section \ref{sec.BP2} presents
the proofs for lattice trees, as well as the minor modifications needed for lattice animals.
Section \ref{sec.SAW2} presents the proofs for self-avoiding walks.

\section{Results}
     \label{sec.results}
     
\subsection{Basic Background and Notation}
   \label{sec.defs}

We denote the standard basis of $\mathbb{R}^d$ by $u^{(1)},\ldots,u^{(d)}$; that is, $u^{(i)}$ is the 
unit vector in the $+x_i$ direction.

We write $\mathbb{Z}^d$ for
the set of points $(x_1,\ldots,x_d)$ in  $\mathbb{R}^d$ whose coordinates $x_i$ are all integers.
The $d$-dimensional hypercubic lattice $\mathbb{L}^d$ 
is the infinite graph embedded in $\mathbb{R}^d$, 
whose sites are the points of $\mathbb{Z}^d$ and whose edges join each pair of sites that
are distance 1 apart.
Let $\mathbb{L}_+^d$ be the part of $\mathbb{L}^d$ that lies in the half-space $\{x:x_1\geq 0\}$.

If $A\subset \mathbb{R}^d$ (or if $A$ is a subgraph of $\mathbb{L}^d$)
and $x\in \mathbb{Z}^d$, then the translation of $A$ by the vector $x$ is denoted $A+x$.

For a subgraph $\rho$ of $\mathbb{L}^d$, let $\mathcal{H}(\rho)$ be the set of sites $x$ of $\rho$
such that $x_1=\rho$.  Thus, referring to Equation (\ref{eq.Zgen}), the quantity
$\sigma(\rho)$ equals $|\mathcal{H}(\rho)|$, the cardinality of $\mathcal{H}(\rho)$.

We shall frequently use superscripts $+$ and $P$ to denote
impenetrable and penetrable surfaces respectively.  Also, we shall use $T$, $A$, and $W$ 
superscripts to denote trees, animals, and (self-avoiding) walks.

\subsection{Branched Polymers:  Trees and animals}
\label{sec.BP1}

A lattice animal is a finite connected subgraph of $\mathbb{L}^d$, and a lattice tree is a 
lattice animal with no cycles.  Each corresponds to a standard discrete model 
of the configuration of a branched polymer.
Let ${\cal{T}}_N$ be the set of all $N$-site lattice trees that contain the origin.
Let $\bar{\cal{T}}_N$ be the set of $N$-site lattice trees whose lexicographically smallest
site is the origin.  (The elements of $\bar{\cal{T}}_N$ correspond to equivalence classes
of all $N$-site lattice trees up to translation.)
Then $|{\cal{T}}_N|\,=\, N\,|\bar{\cal{T}}_N|$.

Let $t_N=|\bar{\cal{T}}_N|$.  It is well known (Klarner, 1967; Klein, 1981) 
that $t_Nt_M\leq t_{N+M}$
for all $N,M\geq 1$, and that $t_N^{1/N}$ has a finite limit $\lambda_d$ with the 
property that 
\begin{equation}
    \label{eq.tNleq}
       t_N\leq \lambda_d^N  \hspace{5mm} \hbox{for every $N$}.
\end{equation}
The notation and results for lattice animals are exactly analogous:  ${\cal{A}}_N$, 
$\bar{\cal{A}}_N$, $a_N=|\bar{\cal{A}}_N|=|{\cal A}_N|/N$, 
$\lambda_{d,A}:= \lim_{n\rightarrow\infty} a_N^{1/N}$,
and $a_N\leq \lambda_{d,A}^N$.

Let ${\cal{T}}_N^+$ be the set of all trees $\tau\in {\cal{T}}_N$ such that 
$\tau\subset \mathbb{L}^d_+$.
Then for every site $x$ of every tree $\tau$ in ${\cal T}_N^+$, we have $x_1\geq 0$.
Observe that $\bar{\cal{T}}_N\subset {\cal{T}}_N^+\subset {\cal{T}}_N$.

We now consider the ensemble of lattice trees in the half-space $\mathbb{L}^d_+$ in which
each site in the boundary plane $x_1=0$ receives unit energy reward.
For real $\beta$, define the partition function
\begin{equation}
   \label{eq.Zdef}
      Z^{T+}_N(\beta)  \;  := \; \sum_{\tau\in {\cal{T}}_N^+} \exp(\beta |\Left(\tau)|)  \,.
\end{equation}
As shown in Theorem 6.23 of Janse van Rensburg (2000),  a concatenation argument
can be used to prove that the limiting free energy
\begin{equation}
  \label{eq.defF}
      {\cal F}^{T+}(\beta)  \;:=  \; \lim_{N\rightarrow\infty}\frac{1}{N}\,\log Z^{T+}_N(\beta)
\end{equation}
exists and is finite for every real $\beta$.  

It is not hard to see that  
the number of trees $\tau$ in ${\cal T}_N^+$ with $|{\cal H}(\tau)|=1$ is exactly
$|{\cal T}_{N-1}^+|$ 
for every $N$, and hence
\begin{equation}
   \label{eq.tN1ZN}
            t_{N-1}e^{\beta}  \; \leq \;  Z^{T+}_N(\beta) \,.
\end{equation}
For $\beta\leq 0$, we also have $Z^+_N(\beta)\leq |{\cal T}_N^+| \leq N t_N$, 
and combining this with Equation (\ref{eq.tN1ZN}) shows that 
\begin{equation}
   \label{eq.Zlimneg}
      {\cal F}^{T+}(\beta)
      \;=\; \log \lambda_d \hspace{5mm}
       \hbox{for every $\beta\leq 0$}.
\end{equation}
This says that the polymer desorbs from the surface whenever $\beta$ is nonpositive---that is,
we have $\beta_c^{T+}\geq 0$.
The following result tells us that, in fact, that the polymer desorbs 
whenever $\beta \leq \lambda_d^{-1}$.

\begin{thm}
\label{prop.tree}  
For lattice trees, 
we have ${\cal F}^{T+}(\beta)\,=\,\log \lambda_d$ for every $\beta \leq \lambda_d^{-1}$.
\end{thm}
Theorem \ref{prop.tree}  says that for adsorption of lattice trees to an impenetrable surface, the
critical point satisfies  $\beta^{T+}_c\geq \lambda_d^{-1}$.  This result is somewhat
better than the 
bound $\beta_c^{T+}\geq \beta^{T+}_c-\beta_c^{TP}\geq \frac{1}{2}\log(1+\lambda_d^{-1})$ 
that follows from
Theorem 4.7 of Janse van Rensburg and You (1998) (which however applies to a larger class of
tree models).  However, the main contribution of our Theorem \ref{prop.tree} is the new method of
proof, rather than the improved numerical value of the bound.

\smallskip

We now consider adsorption at a penetrable surface, and
the relevant ensemble ${\cal T}_N$ of all $N$-site trees that contain the origin.
The corresponding partition function is 
\begin{equation}
   \label{eq.ZPdef}
      Z^{TP}_N(\beta)  \;  := \; \sum_{\tau\in {\cal{T}}_N} \exp(\beta |{\cal H}(\tau)|)  \,.
\end{equation}
As in the impenetrable case, a concatenation argument (see Theorem 6.23 of 
Janse van Rensburg 2000) shows  that the limit
\begin{equation}
  \label{eq.defFP}
      {\cal F}^{TP}(\beta)  \;:=  \; \lim_{N\rightarrow\infty}\frac{1}{N}\,\log Z^{TP}_N(\beta)
\end{equation}
exists and is finite for every real $\beta$.  As was the case for ${\cal F}^{T+}$, 
\begin{equation}
   \label{eq.ZPlimneg}
      {\cal F}^{TP}(\beta)
      \;=\; \log \lambda_d \hspace{5mm}
       \hbox{for every $\beta\leq 0$}.
\end{equation}
It is not hard to show that $0\leq \beta_c^{TP}\leq \beta_c^{T+}\leq \ln(\lambda_d/\lambda_{d-1})$
(see Hammersley et al., 1982, or Janse van Rensburg and You, 1998).
However, in marked contrast to the situation for ${\cal F}^{T+}$, it is generally
believed that ${\cal F}^{TP}(\beta)\,>\, \log \lambda_d$ for every $\beta>0$ ---
i.e., that $\beta^{TP}_c=0$.
Proving this is a challenging open problem.  We shall show that it is a consequence of a 
different
property that has not been proven rigorously but is widely believed to be true.

In the following, we let $\Pr_A$ denote the uniform probability distribution
on the set $A$.   Define the $x_1$-span of a tree $\tau$
to be the number of integers $j$ such that $\tau$ contains a site $v$
with $v_1=j$.  We write $\textrm{Span}(\tau)$ to denote the $x_1$-span of $\tau$.
Since trees are connected, we have
\[    \textrm{Span}(\tau)  \;:=\;  1\,+\, \max\{  |u_1-v_1|  \,: u,v\in \tau\} \,.    \]

\begin{thm}
   \label{prop.treeperm}
Assume there exists $\delta \in (0,\infty)$ such that 
\begin{equation}
   \label{eq.spancond}
    \Pr\!{}_{{\cal T}_N}\left(\left\{ \tau : \,\textrm{Span}(\tau)   \,\leq \, \frac{N}{\log^2 N} \right\} \right)
        \;\geq \; \frac{1}{N^{\delta}}  
\end{equation}
for all sufficiently large $N$.  Then ${\cal F}^{TP}(\beta)\,>\, \log \lambda_d$ for every $\beta>0$
(that is, $\beta_c^{TP}=0$).
\end{thm}

\begin{rem}
   \label{rem.tree}
(\textit{i})
It is generally believed that the expected value of $\textrm{Span}(\tau)$ over ${\cal T}_N$
scales as $N^{\nu}$ for some (dimension-dependent) critical exponent $\nu<1$ (e.g.\ see
section 9.2 of Vanderzande 1998).  This would imply the truth of Equation (\ref{eq.spancond}); 
indeed, it would imply that the left-hand side of (\ref{eq.spancond})
converges to 1 as $N$ tends to $\infty$. 
\\
(\textit{ii}) It will be seen from the proof that the statement of Theorem \ref{prop.treeperm} 
can be strengthened 
slightly, e.g.\ by replacing the square (of the logarithm) by a power greater than 1.
\\
(\textit{iii}) The direct analogues of Theorems \ref{prop.tree} and \ref{prop.treeperm} also hold
for lattice animals   (see Remarks  \ref{rem.treeimp} and  \ref{rem.treeperm}).
\\
(\textit{iv})  There are other ways to define the span of  a tree, but the choice of method
will not substantially affect the statement of the theorem.  Our choice, using the $x_1$ 
coordinate,  is for convenience.
\end{rem}

\subsection{Linear polymers:  Self-avoiding walks}
  \label{sec.SAW1}

An $N$-step self-avoiding walk (SAW) in $\mathbb{L}^d$ is a sequence 
$\omega=(\omega(0),\omega(1),\ldots,\omega(N))$ 
of $N+1$ distinct points of $\mathbb{Z}^d$ such that $\omega(i)$ is a nearest 
neighbour of  $\omega(i-1)$ for $i=1,\ldots,N$.  We write $\omega_j(i)$ to denote the
$j^{th}$ coordinate of the $i^{th}$ point of $\omega$.  The self-avoiding walk is a 
classical model of the configuration of a linear polymer. 

Let ${\cal S}_N$ be the set of all $N$-step self-avoiding walks in $\mathbb{L}^d$ that start
at the origin, and let $c_N=|{\cal S}_N|$.
Then the limit $\mu_d=\lim_{N\rightarrow\infty}c_N^{1/N}$ exists (Hammersley and Morton 1954; 
or see Section 1.2 of Madras and Slade 1993).

Our notation for SAWs is very similar to our notation for trees. 
Let ${\cal S}_N^+$ be the set of all SAWs in ${\cal S}_N$ that are contained in $\mathbb{L}^d_+$.
Then $|{\cal S}_N^+|^{1/N}$ 
also converges to $\mu_d$ (e.g., by Corollary 3.1.6 of Madras and Slade 1993).
The partition function for adsorption at an impenetrable surface is defined to be
\begin{equation}
      Z^{W+}_N(\beta)  \; :=\;  \sum_{\omega\in {\cal S}_N^+} \exp(\beta|\Left(\omega)|)  \,.
               \label{eq.ZNsaw}
\end{equation}
Hammersley et al.\  (1982) proved the existence of the limit
\begin{equation}
  \label{eq.defFwalk}
      {\cal F}^{W+}(\beta)  \;:=  \; \lim_{N\rightarrow\infty}\frac{1}{N}\,\log Z^{W+}_N(\beta)
\end{equation}
for every real $\beta$. 
The following result is the analogue of Theorem \ref{prop.tree} for SAWs, proving
that $\beta^{W+}_c\geq \frac{1}{2}\mu_d^{-2}$.

\begin{thm}
\label{prop.sawimp}
We have ${\cal F}^{W+}(\beta)\,=\,\log \mu_d$ for every $\beta \leq \frac{1}{2}\mu_d^{-2}$.
\end{thm}

For the case of a penetrable surface, let 
\begin{equation}
   \label{eq.ZWPdef}
      Z^{WP}_N(\beta)  \;  := \; \sum_{\tau\in {\cal{S}}_N} \exp(\beta |{\cal H}(\tau)|)  \,.
\end{equation}
Hammersley et al.\  (1982) proved that the limit
\begin{equation}
  \label{eq.defFWP}
      {\cal F}^{WP}(\beta)  \;:=  \; \lim_{N\rightarrow\infty}\frac{1}{N}\,\log Z^{WP}_N(\beta)
\end{equation}
exists and is finite for every real $\beta$, and equals $\log\mu_d$ whenever $\beta\leq 0$.

We define the $x_1$-span of a SAW exactly as for trees:
\[    \textrm{Span}(\omega)  \;:=\;  1\,+\, \max\{  |\omega_1(i)-\omega_1(j)|  \,: 0\leq i,j\leq N\} \,.    \]
We define an $N$-step bridge to be an $N$-step self-avoiding walk with the property that
\[    \omega_d(0)<\omega_d(i)\leq \omega_d(N)    \hspace{5mm}\hbox{for $i=1,\ldots,N$.}
\]
Let ${\cal S}_N^B$ be the set of all bridges in ${\cal S}_N$, and let $b_N=|{\cal S}^B_N|$.
The following result provides a sufficient condition for $\beta^{WP}_c$ to be zero, 
analogously to Theorem \ref{prop.treeperm}.

\begin{thm}
   \label{prop.sawperm}
Assume there exists $\delta \in (0,\infty)$ such that 
\begin{equation}
   \label{eq.Wspancond}
      \Pr\!{}_{{\cal S}^B_N}\left(\left\{ \omega : \,\textrm{Span}(\omega)  \,\leq \, 
      \frac{N}{\log^2 N} \right\} \right)
        \;\geq \; \frac{1}{N^{\delta}}  
\end{equation}
for all sufficiently large $N$.  Then ${\cal F}^{WP}(\beta)\,>\, \log \mu_d$ for every $\beta>0$.
\end{thm}
Similarly to Remark \ref{rem.tree}(\textit{i}), it is generally believed that the left side
of Equation (\ref{eq.Wspancond}) converges to 1 as $N$ tends to infinity.

\section{Branched Polymers: Proofs}
   \label{sec.BP2}

\subsection{Branched Polymers at an Impenetrable Boundary}
  \label{sec.treeimp}

\begin{rem}  
  \label{rem.treeimp}
Everything in this subsection holds if lattice trees are replaced by lattice animals.
\end{rem}


For $\tau\in {\cal{T}}_N^+$, we 
think of the set of sites ${\cal H}(\tau)$ as 
the ``left side of $\tau$''.   The set $\Left(\tau)$ is not empty 
because $\tau$ contains the origin.  For $1\leq k\leq N$, let 
\[    \textrm{left}_N(k)  \;=\; \left|\{ \tau\in {\cal T}^+_N\,: \, |\Left(\tau)|=k\, \}\right|  \,.
\]
%
Then we can write (recalling 
 Equation (\ref{eq.Zdef}))
\begin{equation}
   \label{eq.Zdef2}
     |{\cal T}^+_N| \;=\; \sum_{k=1}^N \textrm{left}_N(k)   
          \hspace{5mm}  \hbox{and} \hspace{5mm}   
      Z^{T+}_N(\beta)  
            \;=\; \sum_{k=1}^N \textrm{left}_N(k)\, e^{\beta k}\,. 
\end{equation}

\noindent
\textbf{Proof of Theorem \ref{prop.tree} :}
Fix $\beta$ such that $0<\beta<\lambda_d^{-1}$.
From Equation (\ref{eq.Zdef2}) we have
\begin{equation}
   \label{eq.ZkNexp}
     Z^{T+}_N(\beta)  
       \;=\;  \sum_{k=1}^{N} \, \sum_{j=0}^{\infty} \, \frac{\beta^jk^j}{j!} \,\textrm{left}_N(k)  \,.
\end{equation}
For any $j\geq 0$ and $k\geq 1$, we have
\begin{equation}
   \label{eq.combin}
      \frac{k^j}{j!} \; \leq \;  \binom{k+j-1}{j}  \,.
\end{equation}
The right hand side of inequality (\ref{eq.combin}) is the number of ways to put $j$ identical balls
into $k$ distinct boxes.  More formally, it is the number of $k$-tuples $(w_1,\ldots,w_k)$ 
of nonnegative integers such that $w_1+\cdots+w_k=j$.  

We shall define a \textit{marked tree} (with $N$ sites) to be a tree $\tau$ in ${\cal{T}}_N^+$ 
that has a nonnegative integer $w(\tau;v)$ assigned to each site $v$ of $\Left(\tau)$.  
(We think of $w(\tau;v)$ as the number of ``marks'' on the site $v$ of $\tau$.)
Let ${\cal T}_N^{(j)}$ be the set of all marked trees $\tau$ with $N$ sites such that the total
number of marks on the sites of $\tau$ is $j$ (that is, $\sum_{v\in \Left(\tau)}w(\tau;v)=j$).
See Figure \ref{fig1}.
Then 
\begin{equation}
   \label{eq.taumark}
      \left| {\cal T}_N^{(j)} \right|  \;=\;  \sum_{k=1}^N  \binom{k+j-1}{j}  \,  \textrm{left}_N(k)  \,.      
\end{equation}      

\setlength{\unitlength}{.45mm}
\begin{figure}
\newsavebox{\dink}
\savebox{\dink}(0,0){\line(0,1){0.8}}
\begin{center}
\begin{picture}(190,90)(0,0)
\put(0,0){
\begin{picture}(70,80)(0,0)
\put(30,30){\line(1,0){10}}
\put(40,20){\line(0,1){10}}
\put(10,20){\line(1,0){40}}
\put(20,10){\line(0,1){10}}
\put(40,50){\line(0,1){10}}
\put(10,40){\line(1,0){20}}
\put(20,60){\line(1,0){20}}
\put(10,20){\line(0,1){20}}
\put(40,50){\line(1,0){30}}
\put(50,10){\line(0,1){10}}
\put(20,40){\line(0,1){20}}
\put(10,70){\line(1,0){20}}
\put(30,60){\line(0,1){10}}
\put(60,30){\line(0,1){30}}
\put(50,40){\line(1,0){10}}
\put(10.1,70){\circle*{1.7}}
\put(10.1,40){\circle*{1.7}}
\put(10.1,30){\circle*{1.7}}
\put(10.1,20){\circle*{1.7}}
\put(12,29.5){\tiny{0}}
\multiput(9,-1)(0,2){42}{\usebox{\dink}}
\multiput(11,-1)(0,2){42}{\usebox{\dink}}
\end{picture}
}
\put(120,0){
\begin{picture}(70,90)(0,0)
\put(30,30){\line(1,0){10}}
\put(40,20){\line(0,1){10}}
\put(10,20){\line(1,0){40}}
\put(20,10){\line(0,1){10}}
\put(40,50){\line(0,1){10}}
\put(10,40){\line(1,0){20}}
\put(20,60){\line(1,0){20}}
\put(10,20){\line(0,1){20}}
\put(40,50){\line(1,0){30}}
\put(50,10){\line(0,1){10}}
\put(20,40){\line(0,1){20}}
\put(10,70){\line(1,0){20}}
\put(30,60){\line(0,1){10}}
\put(60,30){\line(0,1){30}}
\put(50,40){\line(1,0){10}}
\put(10.1,70){\circle*{1.7}}
\put(10.1,40){\circle*{1.7}}
\put(10.1,30){\circle*{1.7}}
\put(10.1,20){\circle*{1.7}}
\put(3,67){3}
\put(3,38){0}
\put(3,28){5}
\put(3,18){3}
\put(12,29.5){\tiny{0}}
\multiput(9,-1)(0,2){42}{\usebox{\dink}}
\multiput(11,-1)(0,2){42}{\usebox{\dink}}
\put(-10,88){number of marks}
\put(5,86){\vector(0,-1){10}}
\end{picture}
}
\end{picture}
\end{center}
\caption{\label{fig1} \textit{Left:}  A tree $\tilde{\tau}$ in ${\cal T}^+_{28}$.  
The vertical dashed double line 
denotes the surface $\{x_1=0\}$.  Here, $|{\cal H}(\tilde{\tau})|=4$.   
\textit{Right:}  A marked tree $\tilde{\tau}$ in ${\cal T}_{28}^{(11)}$. The numbers show the values
of $w(\tilde{\tau};v)$ for each site $v$ in $\Left(\tilde{\tau})$. 
}
\end{figure}
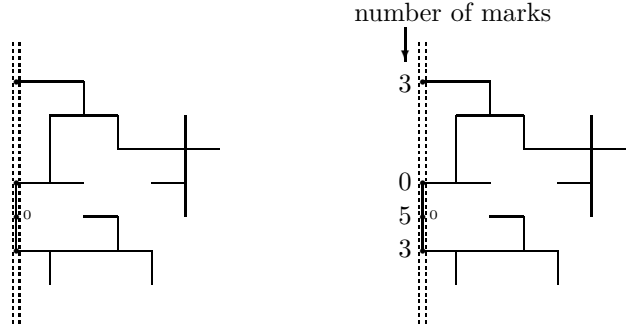
Combining Equations (\ref{eq.ZkNexp}--\ref{eq.taumark}) shows that 
\begin{equation}
    \label{eq.ZbdTj}
      Z^{T+}_N(\beta)  \;\leq \;  \sum_{j=0}^{\infty} \beta^j  \,\left| {\cal T}_N^{(j)} \right|   \,.
\end{equation}      
Now, consider an arbitrary marked tree $\tau\in  {\cal T}_N^{(j)}$.  
For every site $v$ in $\Left(\tau)$, enlarge the tree by attaching  a segment of length 
$w(\tau;v)$ from $v$ to $v-w(\tau;v)u^{(1)}$.
The result is a tree $f(\tau)$ in ${\cal{T}}_{N+j}$ (with no marks).  See Figure \ref{fig2}.
\setlength{\unitlength}{.45mm}
\begin{figure}[h]
\savebox{\dink}(0,0){\line(0,1){0.8}}
\begin{center}
\begin{picture}(210,85)(0,0)
\put(0,0){
\begin{picture}(70,80)(0,0)
\put(30,30){\line(1,0){10}}
\put(40,20){\line(0,1){10}}
\put(10,20){\line(1,0){40}}
\put(20,10){\line(0,1){10}}
\put(40,50){\line(0,1){10}}
\put(10,40){\line(1,0){20}}
\put(20,60){\line(1,0){20}}
\put(10,20){\line(0,1){20}}
\put(40,50){\line(1,0){30}}
\put(50,10){\line(0,1){10}}
\put(20,40){\line(0,1){20}}
\put(10,70){\line(1,0){20}}
\put(30,60){\line(0,1){10}}
\put(60,30){\line(0,1){30}}
\put(50,40){\line(1,0){10}}
\put(10.1,70){\circle*{1.7}}
\put(10.1,40){\circle*{1.7}}
\put(10.1,30){\circle*{1.7}}
\put(10.1,20){\circle*{1.7}}
\put(3,67){3}
\put(3,38){0}
\put(3,28){5}
\put(3,18){3}
\put(12,29.5){\tiny{0}}
\multiput(9,-1)(0,2){42}{\usebox{\dink}}
\multiput(11,-1)(0,2){42}{\usebox{\dink}}
\end{picture}
}
\put(140,0){
\begin{picture}(70,80)(0,0)
\put(30,30){\line(1,0){10}}
\put(40,20){\line(0,1){10}}
\put(10,20){\line(1,0){40}}
\put(20,10){\line(0,1){10}}
\put(40,50){\line(0,1){10}}
\put(10,40){\line(1,0){20}}
\put(20,60){\line(1,0){20}}
\put(10,20){\line(0,1){20}}
\put(40,50){\line(1,0){30}}
\put(50,10){\line(0,1){10}}
\put(20,40){\line(0,1){20}}
\put(10,70){\line(1,0){20}}
\put(30,60){\line(0,1){10}}
\put(60,30){\line(0,1){30}}
\put(50,40){\line(1,0){10}}
\multiput(-19.5,70)(10,0){3}{\circle*{1.3}}
\multiput(-39.5,30)(10,0){5}{\circle*{1.3}}
\multiput(-19.5,20)(10,0){3}{\circle*{1.3}}
\put(-20,70){\line(1,0){30}}
\put(-40,30){\line(1,0){50}}
\put(-20,20){\line(1,0){30}}
\put(10.1,70){\circle*{1.7}}
\put(10.1,40){\circle*{1.7}}
\put(10.1,30){\circle*{1.7}}
\put(10.1,20){\circle*{1.7}}
\put(12,29.5){\tiny{0}}
\multiput(9,-1)(0,2){42}{\usebox{\dink}}
\multiput(11,-1)(0,2){42}{\usebox{\dink}}
\end{picture}
}
\put(80,40){$\lhook\joinrel\relbar\joinrel\rightarrow$}       
\end{picture}
\end{center}
\caption{\label{fig2}  \textit{Left:}  A marked tree $\tau$ in ${\cal T}_{28}^{(11)}$ (see 
  Figure \ref{fig1}).  
   \textit{Right:}  The tree $f(\tau)$ in ${\cal T}_{39}$.}
\end{figure}
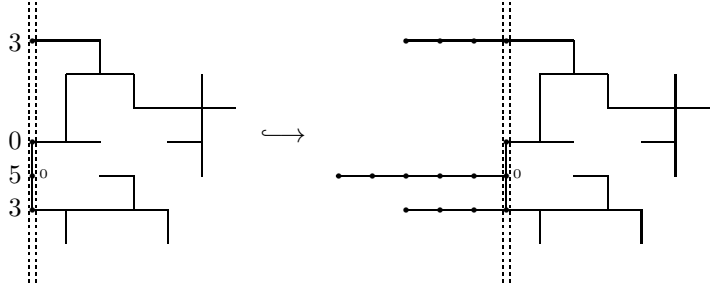
The mapping $f: {\cal T}_N^{(j)} \rightarrow {\cal{T}}_{N+j}$ is 
clearly one-to-one (since $\tau\,=\,f(\tau) \cap \mathbb{L}^d_+$ and the marks are easily
recovered from the segments of $f(\tau)$ outside of $\mathbb{L}^d_+$), and hence
$| {\cal T}_N^{(j)} | \,\leq \,|{\cal T}_{N+j}|\,=\, (N+j)\,t_{N+j}$.      
Combining this with Equations (\ref{eq.ZbdTj}) and (\ref{eq.tNleq}) gives
\begin{eqnarray}
     \nonumber
     Z^{T+}_N(\beta) \; \;\leq \;\; \sum_{j=0}^{\infty} (N+j) \beta^j \lambda_d^{N+j}  
        & = &  \frac{N \,\lambda_d^N}{1-\beta\lambda_d} \,+\,
             \frac{\lambda_d^N\,(\beta \lambda_d)}{(1-\beta\lambda_d)^2}    \\
   \label{eq.Zbdgeom}
       & \leq &    \frac{N \,\lambda_d^N}{(1-\beta\lambda_d)^2}
\end{eqnarray}
(the above series converge because $0<\beta<\lambda_d^{-1}$).  
Equations (\ref{eq.Zbdgeom}) and (\ref{eq.tN1ZN}) imply that 
${\cal F}^{T+}(\beta)\,=\,\log \lambda_d$.

This proves that Equation (\ref{eq.Zlimneg}) extends to every $\beta<\lambda_d^{-1}$.  
The extension to $\beta=\lambda_d^{-1}$ holds by continuity of ${\cal F}^{T+}$ (see
Equation (\ref{eq.Fgen}) and the comments below it).
\hfill$\Box$

\subsection{Branched Polymers at a Penetrable Boundary}
   \label{sec.treeperm}
   
\medskip
\noindent
\textbf{Proof of Theorem \ref{prop.treeperm}:}
A mean-field bound due to Bovier,  Fr\"{o}hlich, and Glaus (1986) (see Section 7.2 of
Slade 2006 for a more detailed proof) says that there exists a constant
$A$ such that 
\begin{equation}
   \label{eq.mfbound}
        1+\sum_{N=1}^{\infty}N^2t_Nz^N  \;\geq \;  \frac{A}{\sqrt{1-\lambda_dz}}
        \hspace{5mm}\hbox{for all $z\in [0,\lambda_d^{-1})$.}
\end{equation}
In particular, the power series on the left diverges at $z=1/\lambda_d$.
It follows that 
\begin{equation}
    \label{eq.iobound}
         t_n   \;\geq \; n^{-4}\lambda_d^n     \hspace{5mm}\hbox{for infinitely many values of $n$}.
\end{equation}                 
Let ${\cal B}_N$ be the set of trees in $\bar{\cal T}_N$ whose $x_1$-span is at most
$N/\log^2N$.    Observe that the left-hand side of Equation (\ref{eq.spancond}) does not
change if we replace $\Pr_{{\cal T}_N}$ by $\Pr_{\bar{\cal T}_N}$.
Thus Equation (\ref{eq.spancond}) says that  $|{\cal B}_N|\,\geq \, t_N/N^{\delta}$.   
By Equation (\ref{eq.iobound}), we obtain  
\begin{equation}
    \label{eq.iobound2}
         |{\cal B}_n|   \;\geq \; n^{-(4+\delta)} \lambda_d^n 
            \hspace{5mm}\hbox{for infinitely many values of $n$}.
\end{equation}                       

For every $N>1$, let 
\begin{equation*}
     {\cal T}^*_N  \;:=\; \left\{ \tau\in {\cal T}_N:  \hbox{0 is the lexicographically smallest site of 
         ${\cal H}(\tau)$}\,  \right\} 
 \end{equation*}
 and 
 \[
     {\cal D}_N  \;:=\;  \{\tau\in {\cal T}^*_N: |{\cal H}(\tau)|\geq \log^2N \}  \,.
\]   
Consider an arbitrary $\tau$ in ${\cal B}_N$.  There must be some integer 
$j\in [0,(N/\log^2N)-1]$ such that
$\tau$ has at least $\log ^2N$ sites $x$ satisfying $x_1=j$.   
Let $\hat{x}$ be the lexicographically smallest site in  $\{x\in \tau: x_1=j\}$, 
and let $\hat{\tau}$ be the translation of $\tau$ by the vector $-\hat{x}$.
Then  $\hat{\tau}\in {\cal D}_N$.   
Observe that each $\hat{\tau}$ uniquely determines $\tau$, since ${\cal B}_N\subset \bar{\cal T}_N$
and no two trees in $\bar{\cal T}_N$ can be translations of one another.
Therefore 
\begin{equation}
  \label{eq.DNBN}
     |{\cal D}_N|\,\geq \,|{\cal B}_N|.
\end{equation}

Now fix $\beta>0$.  By Equation (\ref{eq.iobound2}), there exists an integer $n$ for which 
\begin{equation}
    \label{eq.fixn}
     |{\cal B}_n|\exp(\beta \log^2 n) \,>\,\lambda_d^n.  
\end{equation}
Fix this $n$ for the rest of the proof.  

We can concatenate members of ${\cal D}_n$ by translating them along vectors in
the hyperplane $x_1=0$.  Details are given in Section \ref{sec.cat} below.
For any integer $k\geq 2$, we can concatenate any $k$ members of ${\cal D}_n$ in this way to 
produce a member $\tilde{\tau}$ of ${\cal T}_{kn}$ with ${\cal H}(\tilde{\tau})\,\geq \, k\log^2n$.
Moreover, this map $({\cal D}_n)^k\rightarrow {\cal T}_{kn}$ is injective (see Section \ref{sec.cat}).
Therefore, using Equation (\ref{eq.DNBN}), we have 
\begin{align}
      Z_{kn}^{TH}(\beta)  \;& \geq \;  |{\cal D}_n|^k \exp(\beta k \log^2n)
        \nonumber  \\
           & \geq \;   |{\cal B}_n|^k \exp(\beta k \log^2n)   
           \hspace{5mm}(k=1,2,\ldots).
        \label{eq.ZPconcat}
\end{align}
Take the $(kn)^{th}$ root of Equation (\ref{eq.ZPconcat}) and let $k\rightarrow\infty$.
Since the limit of the left-hand side exists, we obtain
\[
    \exp({\cal F}^{TH}(\beta) ) \;\geq \;   \left( |{\cal B}_n| \exp(\beta \log^2n) \right)^{1/n},
\]
and the right hand side is strictly greater than $\lambda_d$ by Equation (\ref{eq.fixn}).
This proves that ${\cal F}^{TH}(\beta) \,>\, \log \lambda_d$. 
\hfill  $\Box$    

\begin{rem}  
   \label{rem.treeperm}
The analogue of Equation (\ref{eq.mfbound}) for lattice animals appears in 
Section 1.3 of Hara and Slade (1990).  Everything else in this section extends
immediately to lattice animals.
\end{rem}

\subsection{Concatenation of Lattice Branched Polymers}
    \label{sec.cat}
    
This section describes a concatenation procedure that preserves the number of
sites in the surface $x_1=0$.    We shall discuss trees, but the argument for
animals is essentially the same.

Let $N$ and $M$ be positive integers.  We shall describe an operation $\oplus$
such that, for every pair of trees $\tau\in{\cal T}^*_N$ and $\psi\in{\cal T}^*_M$,
we obtain a tree $\tau\oplus\psi\in {T}^*_{N+M}$ such that 
$|{\cal H}(\tau\oplus\psi)| \,=\, |{\cal H}(\tau)|\,+\,|{\cal H}(\psi)|$.  Moreover, the operation 
$\oplus: {\cal T}^*_N\times {\cal T}^*_M \rightarrow{\cal T}^*_{N+M}$
is one-to-one.

Let $\tau\in{\cal T}^*_N$ and $\psi\in{\cal T}^*_M$.  Let
\[   K \;=\; \max\left\{ k\in \mathbb{Z}: (\psi+ku^{(2)})\cap \tau \neq \emptyset \right\} \,.
\]
Since $\psi\cap\tau$ contains the origin, we see that $K\geq 0$. 
Let $v$ be a site in $(\psi+Ku^{(2)})\cap \tau$, and let $b$ be the edge from $v$ to $v+u^{(2)}$.  
Observe that $\psi+(K+1)u^{(2)}$ contains $v+u^{(2)}$ but contains no point of $\tau$.  
Therefore $(\psi+(K+1)u^{(2)})\cup \tau \cup b$ is a tree, which we shall call $\theta$.
We define $\tau\oplus\psi$ to be $\theta$.  We shall now check that $\theta$ has
the claimed properties of $\oplus$.

First observe that the construction ensures that we have 
\begin{verse}
   \textbf{Property A:}  \hspace{1mm} ${\cal H}(\theta)$ is the disjoint union of
   and ${\cal H}(\psi)+(K+1)u^{(2)}$ and   ${\cal H}(\tau)$.
\end{verse}
It is clear that $\theta\in {\cal T}_{N+M}$.  To show that  $\theta\in {\cal T}^*_{N+M}$, we
must show that 0 is the lexicographically smallest site of ${\cal H}(\theta)$.  
But this follows from Property A, the fact that 0 is the lexicographically smallest site of 
${\cal H}(\tau)$ and of ${\cal H}(\psi)$, and our earlier observation that $K\geq 0$.
The relation $|{\cal H}(\tau\oplus\psi)| \,=\, |{\cal H}(\tau)|\,+\,|{\cal H}(\psi)|$ also 
follows from Property A.

It remains to show that $\oplus$ is one-to-one, i.e.\ that we can recover $\tau$ and $\psi$
knowing $\theta$ (for given $N$ and $M$).
To do this, we first observe that for the edge $b$ in our construction,  the 
following property holds with $e=b$:
\begin{verse}
   \textbf{Property B:}  \hspace{1mm} Deleting the edge $e$ from $\theta$ creates 
   two components, and the component containing the origin has exactly $N$ sites.
\end{verse}
In general, there may be two or more edges $e$ of $\theta$ that satisfy Property B, so 
we need to decide which of them is $b$.
Let $J=\max\{j\in \mathbb{Z}:  ju^{(2)}\in \theta\}$.  Since $(K+1)u^{(2)}\in \theta$, we see
that $J\geq K+1$.  Thus, whatever $\tau$ and $\psi$ are, we know that $0\in \tau$ 
and $Ju^{(2)}\not\in \tau$ (by the definition of $K$ and the fact that $Ju^{(2)}\in \psi+Ju^{(2)}$).  
Therefore the edge $b$ belongs to $\pi$, where $\pi$ 
is any path in $\theta$ from 0 to $Ju^{(2)}$.  
(When $\theta$ is a  tree, there is only one such path.)
Furthermore, it is not hard to see that at most one edge of $\pi$ can satisfy Property B.
Therefore the edge $b$ is determined from $\theta$, and hence $\tau$ and $\psi$ are determined.
This proves that $\oplus$ is one-to-one.

\section{Linear Polymers}
    \label{sec.SAW2}

\subsection{Self-Avoiding Walks at an Impenetrable Boundary}
   \label{sec.SAWimp}


\noindent
\textbf{Proof of Theorem \ref{prop.sawimp}:}
Hammersley et al.\  (1982) proved that ${\cal F}^{W+}(\beta)=\log\mu_d$ 
for every $\beta\leq 0$, so we shall only consider positive $\beta$.
The general idea of the proof is the same as for trees (Theorem \ref{prop.tree}), 
but there is a technical 
difficulty when it comes to proving the analogue of $|{\cal T}_N^{(j)}| \,\leq |{\cal T}_{N+j}|$.
To get around this, we introduce a slightly different model of adsorption, in which we weight
a walk according the number of edges in the surface.  For $\omega\in{\cal S}_N^+$, 
define $\Left\Left(\omega)$ to be the set of edges of $\omega$ that have both endpoints
in $\{x\in \mathbb{Z}^d:x_1=0\}$, and define
\[
    Z^{WW+}_N(\beta)  \;  :=\;  \sum_{\omega\in {\cal S}_N^+} \exp(\beta|\Left\Left(\omega)|) \,.
\]
Then $|\Left(\omega)|   \;\leq \;   2\,|\Left\Left(\omega)|$
for every $\omega\in {\cal S}_N^+$, 
and hence for every $\beta\geq 0$ we have
\begin{equation}
  \label{eq.Zww}
          \; Z_N^{W+}(\beta)   \;\leq \;  Z_N^{WW+}(2\beta)  \,.
\end{equation}
We define a \textit{marked walk} (with $N$ sites) to be a SAW $\omega$ in ${\cal S}_N^+$ that 
has a nonnegative integer $m(\omega;b)$ assigned to each edge $b$ of $\Left\Left(\omega)$.
Let ${\cal S}_N^{(j)}$ be the set of all marked walks $\omega$ with $N$ sites such that
$\sum_{b\in\Left\Left(\omega)}m(\omega;b)\,=\,j$.  Then the same argument as in the 
proof of Theorem \ref{prop.tree} shows that
\begin{equation}
    \label{eq.ZbdSj}
      Z^{WW+}_N(2\beta)  \;\leq \;  \sum_{j=0}^{\infty} (2\beta)^j  \,\left| {\cal S}_N^{(j)} \right|   \,.
\end{equation}    

Now, fix a positive $\beta< \frac{1}{2}\mu_d^{-2}$.  Choose $\epsilon>0$ small enough so that
$2\beta(\mu_d+\epsilon)^2<1$.  Then there exists a constant $A$ such that
\begin{equation}
   \label{eq.sawsum}
       \sum_{n=0}^Mc_n   \;\leq \;  A(\mu_d+\epsilon)^M   \hspace{5mm}\mbox{for all $M\geq 0$}.
\end{equation}
Consider an arbitrary marked walk $\omega$ in ${\cal S}_N^{(j)}$.  Let $E_1$ be the 
set of edges of $\omega$ that are not in $\Left\Left(\omega)$.  Let $E_2$ be the 
set of edges in $\Left\Left(\omega)$ after each edge is translated in the $-x_1$ 
direction by a distance equal to the number of marks on that edge:
\[   E_2  \;=\; \{b-m(\omega;b)u^{(1)}: b\in \Left\Left(\omega) \}  \,.
\]
Let $f(\omega)$ be the shortest SAW starting at the origin that contains all edges 
of $E_1\cup E_2$ and all of whose remaining edges are parallel to $\pm u^{(1)}$.  
Observe that $f(\omega)$ is obtained by adding at most $2j$ edges
to $E_1\cup E_2$.
It is not hard to see that the function $f:{\cal S}_N^{(j)}\rightarrow \bigcup_{n=N}^{N+2j}{\cal S}_N$ 
is one-to-one, so by Equation (\ref{eq.sawsum})
\[      |{\cal S}_N^{(j)}|   \;\leq \;   A(\mu_d+\epsilon)^{N+2j}.   
\]
From this and Equation (\ref{eq.ZbdSj}), and our choice of $\epsilon$, we obtain
\[
    Z_N^{WW+}(2\beta)  \;\leq\;  \frac{A (\mu_d+\epsilon)^N}{1-2\beta(\mu_d+\epsilon)^2}  \,.
\]
Combining this with Equation (\ref{eq.Zww}) proves that 
${\cal F}^{W+}(\beta) \,\leq\,\log(\mu_d+\epsilon)$.
Since $\epsilon$ can be made arbitrarily small, and since 
${\cal F}^{W+}(\beta)\,\geq \,{\cal F}^{W+}(0)\,=\,\log\mu_d$, we are done.
\hfill $\Box$

\subsection{Self-Avoiding Walks at a Penetrable Boundary}
    \label{sec.sawpen}

\noindent
\textbf{Proof of Theorem \ref{prop.sawperm}:}
First observe that if $\omega\in {\cal S}_N^B$, then $\omega(1)=(0,\ldots,0,1)$ and 
$|\omega_1(N)|\leq N-1$.

It is known   
that the series $\sum_{n=1}^{\infty}b_nz^n$ diverges at $z=\mu_d^{-1}$ 
(Kesten, 1963; or Corollary 3.1.8 of Madras and Slade 1993) .  Therefore we have
\begin{equation}
    \label{eq.iobound3}
        b_n  \;\geq \; n^{-2} \mu_d^n 
            \hspace{5mm}\hbox{for infinitely many values of $n$}.
\end{equation}                       
For every $N>1$, let 
\[   
     {\cal D}_N  \;:=\;  \{\omega\in {\cal S}^B_N: \textrm{Span}(\omega) \,\leq \,N/\log^2N \}  \,.
\]   
By the assumption (\ref{eq.Wspancond}), $|{\cal D}_N|/b_N \,\geq \,N^{-\delta}$ for sufficiently large $N$. Therefore by (\ref{eq.iobound3}),
\begin{equation}
    \label{eq.iobound4}
        |{\cal D}_n|  \;\geq \; n^{-(2+\delta)} \mu_d^n 
            \hspace{5mm}\hbox{for infinitely many values of $n$}.
\end{equation}           

Fix $\beta>0$. 
Fix a positive  integer $n$ such that $\frac{\beta}{2}\log^2n > \log(4n^{4+\delta})$ and 
the inequality of (\ref{eq.iobound4}) holds.
For integers   $j$ and $m$ let
\[ 
    {\cal D}_{n,j,m}  \;:=\;  \{\omega\in {\cal D}_n:   |\{i:\omega_1(i)=j\}|\geq \log^2n,\,
       \omega_1(n)=m\}\,.
\]

Since 
\[  {\cal D}_n\,=\,\bigcup_{j=-(n-1)}^{n-1}\bigcup_{m=-(n-1)}^{n-1}{\cal D}_{n,j,m} 
\] 
and by symmetry,
there exist integers $J\geq 0$ and $M$  such that 
$|{\cal D}_{n,J,M}|\,\geq \,|{\cal D}_n|/(2n-1)^2$.  
By this  and (\ref{eq.iobound4}),
\begin{equation}
   \label{eq.DnJMbd}
    |{\cal D}_{n,J,M}|  
    \;\geq \;  \frac{\mu_d^n}{4n^{4+\delta}} \,.
\end{equation}

For two SAWs $\omega=(\omega(0),\ldots,\omega(N))$ and $\psi=(\psi(0),\ldots,\psi(M))$,
we define the concatenation $\omega\oplus\psi$ to be the $(N+M)$-step walk
$\theta$ defined by
\begin{align*}
     \theta(i)   \; & =\;  \omega(i)  \hspace{25mm} \hbox{for $i=0,\ldots, N$, and }   \\
         \theta(N+j) \;&=\; \omega(N)+\psi(j)-\psi(0)   
            \hspace{5mm} \hbox{for $j=1,\ldots, M$} \,.
\end{align*}                      
In general, $\theta$ need not be self-avoiding.   However, if $\omega$ and $\psi$ are both
bridges, then $\theta$ is self-avoiding---indeed, $\theta$ is a bridge.    Thus
$\oplus$ defines a one-to-one map from ${\cal S}^B_N\times {\cal S}^B_M$ into
${\cal S}^B_{N+M}$.      

Suppose now that $\omega\in {\cal D}_{n,J,M}$ and $\psi\in {\cal D}_{n,-J,-M}$, and
let $\theta\,=\,\omega\oplus\psi$.  Then $\theta$ is a $(2n)$-step bridge
such that $\theta_1(2n)=0$ and $|\{i:\theta_1(i)=J\}|\geq \log^2 n$  (the inequality
is due only to sites in the first half of $\theta$).
We shall use these observations in the construction that follows.

For any positive integer $k$, let $\omega^{[1]},\ldots,\omega^{[k]}$ be bridges in
${\cal D}_{n,J,M}$ and let $\psi^{[1]},\ldots,\psi^{[k]}$ be bridges in ${\cal D}_{n,-J,-M}$.
Consider  the bridge $\pi$ obtained by repeated concatenation of these bridges:
\[
  \pi \;:=\;   \omega^{[1]}\oplus\psi^{[1]} \oplus \omega^{[2]} \oplus \psi^{[2]} \oplus
    \cdots  \oplus \omega^{[k]}\oplus \psi^{[k]} \,.
 \]
Then $|\{i:\pi_1(i)=J\}|\geq k\log^2 n$.  Next,  
let $\xi$ be the $(J+1)$-step bridge with $\xi(0)=0$ and $\xi(J{+}1)=(-J,0,\ldots,0,1)$.
For $\zeta:=\xi\oplus\pi$, we have 
$\zeta\in {\cal S}^B_{J+1+2kn}$ and $|{\cal H}(\zeta)| \,\geq \, k\log^2 n$.  
 Since
 $\zeta$ unambiguously determines the $\omega^{[i]}$'s and $\psi^{[i]}$'s, it follows that
 \begin{align*}
     |\{\zeta \in {\cal S}^B_{J+1+2kn}:  |{\cal H}(\zeta)| \,\geq \, k\log^2 n \}|   
        \; & \geq \;   \left( |{\cal D}_{n,J,M}|\,|{\cal D}_{n,-J,-M}|\right)^k   \\
        & = \;    |{\cal D}_{n,J,M}|^{2k}            \hspace{8mm}\hbox{(by symmetry).}
 \end{align*}  
 Using this and Equation (\ref{eq.DnJMbd}), we see that 
 \[  
     Z^{WP}_{J+1+2kn}(\beta)  \;\geq\; 
       \exp(\beta k\log^2 n) \,\frac{\mu_d^{2kn} }{(4n^{4+\delta})^{2k}} \,.
 \]
 Therefore
 \[
    \frac{\log Z^{WP}_{J+1+2kn}(\beta) }{J+1+2kn}    \;\geq\;  
    \frac{\beta k \log^2n -2k\log(4n^{4+\delta}) +2kn\log\mu_d}{J+1+2kn}   \,.
 \]
 Now let $k\rightarrow\infty$, and we obtain
 \begin{align*}
    {\cal F}^{WP}(\beta)  \; & \geq \; 
     \frac{1}{n} \left( \frac{\beta \log^2 n} {2} \,-\, \log(4n^{n+\delta}) \right)   \,+\, \log \mu_d \\
     & > \;  \log\mu_d \,,
 \end{align*}
where the strict inequality follows from our choice of $n$.  This proves the result.
\hfill $\Box$

\section*{Acknowledgments}

This research was supported in part by a Discovery Grant from the Natural Sciences and
Engineering Research Council of Canada.  Part of this work was done while the author was
visiting the Fields Institute for Research in Mathematical Sciences.

\end{document}